\documentclass[12pt,3p]{elsarticle}

\journal{Physics Letters B}

\begin{document}

\begin{frontmatter}

\title{Hypergravity in $AdS_3$}

\author[ihep]{Yu.M.~Zinoviev\corref{cor1}} 

\cortext[cor1]{Corresponding author}

\address[ihep]{Institute for High Energy Physics,
Protvino, Moscow Region, 142280, Russia}

\begin{abstract}
Thirty years ago Aragone and Deser showed that in three dimensions
there exists a consistent model describing interaction for massless
spin-2 and spin-5/2 fields. It was crucial that these fields lived in
a flat Minkowski space and as a result it was not possible to deform
such model into anti-de Sitter space. In this short note we show that
such deformation becomes  possible provided one compliment to the
model with massless spin-4 field. Resulting theory can be considered
as a Chern-Simons one with a well-known supergroup $OSp(1,4)$.
Moreover, there exists a straightforward generalization to the
$OSp(1,2n)$ case containing a number of bosonic fields with even spins
$2,4,\dots,2n$ and one fermionic field with spin $n+1/2$.
\end{abstract}

\end{frontmatter}

\section{Introduction}

Thirty years ago Aragone and Deser showed \cite{AD84} that in three
dimensions there exists a consistent model describing interaction for
massless spin-2 and spin-$\frac{5}{2}$ fields. It was crucial that
these fields lived in a flat Minkowski space and as a result it was
not possible to deform such model into anti-de Sitter space. Taking
into account crucial role that $AdS$ background plays in all massless
higher spin theories it is natural to look for generalization of such
model admitting deformation into $AdS$ space. In this note we show
that such deformation becomes  possible provided one compliment to a
model with massless spin-4 field. Resulting theory can be considered
as a Chern-Simons one with a well-known supergroup $OSp(1,4)$.
Moreover, there exists a straightforward generalization to the
$OSp(1,2n)$ case containing a number of bosonic fields with even spins
$2,4,\dots,2n$ and one fermionic field with spin $n+1/2$.

The paper is organized as follows. In Section 2 in our current
formalism (see notations and conventions below) we reproduce the 
well-known fact (see e.g. \cite{Wit88}) that three-dimensional gravity
in $AdS$ background can be considered as a Chern-Simons theory with
group $SO(2,1) \otimes SO(2,1)$. Then in Section 3 we start directly
with the Chern-Simons theory with $OSp(1,2)$ supergroup and show that
in $AdS$ background it corresponds to minimal $(1,0)$ supergravity
(for the general case of extended $(M,N)$ supergravities, see
\cite{AT86}). At last, in Section 4 we consider straightforward
generalization of such supergravity model to the well-known $OSp(1,4)$
supergroup and show that in $AdS$ background it describes interacting
system of massless spin-2, spin-4 and spin-$\frac{5}{2}$ fields.

\noindent
{\bf Notations and conventions} We will use a multispinor frame-like
formalism where all gauge fields are one-forms but with all local
indices replaced with the completely symmetric spinor ones. Spinor
indices $\alpha,\beta=1,2$ will be raised and lowered with
$\varepsilon^{\alpha\beta} = - \varepsilon^{\beta\alpha}$ such that
$\varepsilon^{\alpha\beta} \varepsilon_{\beta\gamma} = -
\delta^\alpha{}_\gamma$. For the $AdS_3$ background we will use
background frame $e^{\alpha\beta} = e^{\beta\alpha}$ and $AdS_3$
covariant derivative $D$ normalized so that
$$
D \wedge D \xi^\alpha = \frac{\lambda^2}{4} e^\alpha{}_\gamma \wedge
e^{\beta\gamma} \xi_\beta
$$
In what follows we will not write the symbol $\wedge$ explicitly.

\section{Gravity in $AdS_3$}

In this section we consider frame formulation of gravity in $AdS_3$
background. We need two one-forms $h^{\alpha\beta}$ and
$\omega^{\alpha\beta}$ which are symmetric bi-spinors. The Lagrangian
(being three-form) looks like:
\begin{eqnarray}
{\cal L} &=& \omega_{\alpha\beta} e^\alpha{}_\gamma
\omega^{\beta\gamma} + \omega_{\alpha\beta} D h^{\alpha\beta} +
\frac{\lambda^2}{4} h_{\alpha\beta} e^\alpha{}_\gamma h^{\beta\gamma}
\nonumber \\
 && + a_0 h_\alpha{}^\beta \omega_\beta{}^\gamma
\omega_\gamma{}^\alpha + \frac{a_0\lambda^2}{12} h_\alpha{}^\beta
h_\beta{}^\gamma h_\gamma{}^\alpha
\end{eqnarray}
where $a_0$ is a coupling constant. This Lagrangian is invariant under
the following local gauge transformations:
\begin{eqnarray}
\delta \omega^{\alpha\beta} &=& D \eta^{\alpha\beta} +
\frac{\lambda^2}{4} e^{(\alpha}{}_\gamma \xi^{\beta)\gamma} \nonumber
\\
 && + a_0 \omega^{(\alpha}{}_\gamma \eta^{\beta)\gamma} +
\frac{a_0\lambda^2}{4} h^{(\alpha}{}_\gamma \xi^{\beta)\gamma}
\nonumber \\
\delta h^{\alpha\beta} &=& D \xi^{\alpha\beta} + 
e^{(\alpha}{}_\gamma \eta^{\beta)\gamma} \\
 && + a_0 \omega^{(\alpha}{}_\gamma \xi^{\beta)\gamma} + a_0 
h^{(\alpha}{}_\gamma \eta^{\beta)\gamma} \nonumber
\end{eqnarray}
Now let us introduce new variables:
\begin{equation}
\hat{\omega}^{\alpha\beta} = \omega^{\alpha\beta} + 
\frac{\lambda}{2} h^{\alpha\beta}, \qquad 
\hat{h}^{\alpha\beta} = \omega^{\alpha\beta} - \frac{\lambda}{2}
h^{\alpha\beta}
\end{equation}
and similarly for the parameters of gauge transformations:
\begin{equation}
\hat{\eta}^{\alpha\beta} = \eta^{\alpha\beta} + 
\frac{\lambda}{2} \xi^{\alpha\beta}, \qquad
\hat{\xi}^{\alpha\beta} = \eta^{\alpha\beta} - \frac{\lambda}{2}
\xi^{\alpha\beta}
\end{equation}
In terms of these variables the Lagrangian becomes a sum of two
independent Lagrangians for $\hat{\omega}$ and $\hat{h}$ fields, each
one having its own gauge symmetry. In what follows we restrict
ourselves with one field $\hat{\omega}$ only. Corresponding Lagrangian
has the form:
\begin{equation}
{\cal L} = \frac{1}{2} \hat{\omega}_{\alpha\beta} D
\hat{\omega}^{\alpha\beta} + \frac{\lambda}{2}
\hat{\omega}_{\alpha\beta} e^\alpha{}_\gamma
\hat{\omega}^{\beta\gamma} + \frac{a_0}{3} \hat{\omega}_\alpha{}^\beta
\hat{\omega}_\beta{}^\gamma \hat{\omega}_\gamma{}^\alpha
\end{equation}
and is invariant under the following gauge transformation:
\begin{equation}
\delta \hat{\omega}^{\alpha\beta} = D \hat{\eta}^{\alpha\beta} +
\frac{\lambda}{2} e^{(\alpha}{}_\gamma \hat{\eta}^{\beta)\gamma} + a_0
\hat{\omega}^{(\alpha}{}_\gamma \hat{\eta}^{\beta)\gamma}
\end{equation}
Now let us introduce convenient combination
\begin{equation}
\hat{\omega}_0{}^{\alpha\beta} = \omega_0{}^{\alpha\beta} +
\frac{\lambda}{2} e^{\alpha\beta}
\end{equation}
where $\omega_0{}^{\alpha\beta}$ is an $AdS_3$ background Lorentz
connection, so that
\begin{equation}
d \hat{\omega}_0{}^{\alpha\beta} + \hat{\omega}_0{}^\alpha{}_\gamma
\hat{\omega}_0{}^{\beta\gamma} = 0
\end{equation}
Here and in what follows $d$ is a usual external derivative $d^2 = 0$.
Then introducing new variable 
$$
\Omega^{\alpha\beta} = \frac{1}{a_0} \hat{\omega}_0{}^{\alpha\beta} +
\hat{\omega}^{\alpha\beta}
$$
the Lagrangian, gauge transformations and equations can be rewritten
in a very simple form:
\begin{equation}
{\cal L} = \frac{1}{2} \Omega_{\alpha\beta} d \Omega^{\alpha\beta} +
\frac{a_0}{3} \Omega_\alpha{}^\beta \Omega_\beta{}^\gamma
\Omega_\gamma{}^\alpha 
\end{equation}
\begin{equation}
\delta \Omega^{\alpha\beta} = d \hat{\eta}^{\alpha\beta} + a_0
\Omega^{(\alpha}{}_\gamma \hat{\eta}^{\beta)\gamma}
\end{equation}
\begin{equation}
d \Omega^{\alpha\beta} + a_0 \Omega^\alpha{}_\gamma
\Omega^{\beta\gamma} = 0
\end{equation}
Thus we have a Chern-Simons gauge theory with the $Sp(2) \sim
SO(2,1)$ group. Nice feature of such formulation beyond its simplicity
is its background independence, while $AdS_3$ background appears just
as a particular solution of equations. In the next two sections we
will use the reverse procedure, namely we will start directly with
such background independent formulation and then to see the field
content we will go back to $AdS_3$ background.

\section{$OSp(1,2)$ supergravity}

As is well known \cite{AT86} all $(N,M)$ extended supergravities in
$AdS_3$ can be considered as Chern-Simons theories with the
supergroups $OSp(N,2) \otimes OSp(M,2)$. The simplest example is the
$(1,0)$ supergravity that contains just two massless fields: spin-2
and spin 3/2 ones. Here we reproduce this model to illustrate our
formalism.

Model contains two one-form fields: spin-2 $\Omega^{\alpha\beta}$ and
spin-$\frac{3}{2}$ $\Psi^\alpha$ ones. Let us consider the following
ansatz for the Lagrangian
\begin{eqnarray}
{\cal L} &=& \frac{1}{2} \Omega_{\alpha\beta} d \Omega^{\alpha\beta} +
\frac{i}{2} \Psi_\alpha d \Psi^\alpha \nonumber \\
 && + \frac{a_0}{3} \Omega_\alpha{}^\beta \Omega_\beta{}^\gamma
\Omega_\gamma{}^\alpha + b_0 \Psi_\alpha \Omega^{\alpha\beta}
\Psi_\beta
\end{eqnarray}
and corresponding gauge transformations:
\begin{eqnarray}
\delta \Omega^{\alpha\beta} &=& d \eta^{\alpha\beta} + \alpha_1
\Omega^{(\alpha}{}_\gamma \eta^{\beta)\gamma} + \alpha_2
\Psi^{(\alpha} \xi^{\beta)} \nonumber \\
\delta \Psi^\alpha &=& d \xi^\alpha + \alpha_3 \eta^{\alpha\beta}
\Psi_\beta + \alpha_4 \Omega^{\alpha\beta} \xi_\beta
\end{eqnarray}
Variations of the Lagrangian under the $\eta^{\alpha\beta}$
transformations have the form:
\begin{eqnarray*}
\delta_\eta {\cal L} &=& 2(\alpha_1-a_0) d \Omega_{\alpha\beta}
\Omega^\alpha{}_\gamma \eta^{\beta\gamma} \\
 && + (i\alpha_3+2b_0) d \Psi_\alpha \eta^{\alpha\beta} \psi_\beta \\
 && + 2b_0(\alpha_3-\alpha_1) \Psi_\alpha \Omega^{\alpha\beta}
\eta_\beta{}^\gamma \Psi_\gamma
\end{eqnarray*}
so we have to put
$$
\alpha_1 = \alpha_3 = a_0, \qquad  b_0 = - \frac{ia_0}{2}
$$
At the same time variations under the $\xi^\alpha$ transformations
look like:
\begin{eqnarray*}
\delta_\xi {\cal L} &=& 2(\alpha_2+b_0) d \Omega_{\alpha\beta}
\Psi^\alpha \xi^\beta \\
 && + (i\alpha_4-2b_0) d \Psi_\alpha \Omega^{\alpha\beta} \xi_\beta \\
 &&+ 2(b_0\alpha_4-a_0\alpha_2) \Psi_\alpha \Omega^{\alpha\beta}
\Omega_\beta{}^\gamma \xi_\gamma 
\end{eqnarray*}
so we obtain
$$
\alpha_2 = \frac{ia_0}{2}, \qquad \alpha_4 = - a_0
$$
Thus all coefficients in the Lagrangian and gauge transformations
are expressed in terms of just one coupling constant $a_0$. Now let us
consider equations that follow from this Lagrangian:
\begin{eqnarray}
 && d \Omega^{\alpha\beta} + a_0 \Omega^\alpha{}_\gamma
\Omega^{\beta\gamma} + \frac{ia_0}{2} \Psi^\alpha \Psi^\beta = 0
\nonumber \\
 && d \Psi^\alpha + a_0 \Omega^\alpha{}_\beta \Psi^\beta = 0
\end{eqnarray}
It is easy to see that $AdS_3$ background is a solution of these
equations, namely:
$$
\Omega_0{}^{\alpha\beta} = \frac{1}{a_0}
\hat{\omega}_0{}^{\alpha\beta}, \qquad \Psi_0{}^\alpha = 0
$$
Again using $\Omega^{\alpha\beta} = \frac{1}{a_0}
\hat{\omega}_0{}^{\alpha\beta} + \omega^{\alpha\beta}$ we obtain the
Lagrangian for $(1,0)$ supergravity in $AdS_3$ background:
$$
{\cal L} = {\cal L}_0 + {\cal L}_1
$$
\begin{eqnarray}
{\cal L}_0 &=& \frac{1}{2} \omega_{\alpha\beta} D \omega^{\alpha\beta}
+ \frac{\lambda}{2} \omega_{\alpha\beta} e^\alpha{}_\gamma
\omega^{\beta\gamma} \nonumber \\
 && + \frac{i}{2} \Psi_\alpha D \Psi^\alpha + \frac{i\lambda}{4}
\Psi_\alpha e^\alpha{}_\beta \Psi^\beta \\
{\cal L}_1 &=& \frac{a_0}{3} \omega_\alpha{}^\beta
\omega_\beta{}^\gamma \omega_\gamma^\alpha - \frac{ia_0}{2}
\Psi_\alpha \omega^{\alpha\beta} \Psi_\beta
\end{eqnarray}
where ${\cal L}_0$ is just the sum of the free Lagrangians for
massless spin-2 and spin-$\frac{3}{2}$ fields, while ${\cal L}_1$
describes self-interaction of graviton and its interaction with
gravitino. Corresponding gauge transformations take the form:
\begin{eqnarray}
\delta \omega^{\alpha\beta} &=& D \eta^{\alpha\beta} +
\frac{\lambda}{2} e^{(\alpha}{}_\gamma \eta^{\beta)\gamma} \nonumber
\\
 && + a_0 \omega^{(\alpha}{}_\gamma \eta^{\beta)\gamma} + 
\frac{ia_0}{2} \Psi^{(\alpha} \xi^{\alpha)} \nonumber \\
\delta \Psi^\alpha &=& D \xi^\alpha + \frac{\lambda}{2}
e^\alpha{}_\beta \xi^\beta \\
 && + a_0 \eta^{\alpha\beta} \Psi_\beta + a_0 \omega^\alpha{}_\beta
\xi^\beta \nonumber
\end{eqnarray}

\section{$OSp(1,4)$ hypergravity}

One of the important and peculiar features of three-dimensional higher
spin theories is that corresponding infinite dimensional
(super)algebras admit finite dimensional truncations \cite{Vas91}.
For example, in \cite{CFPT10} it was shown that consistent models
containing massless fields with integer spins $2,3,\dots,N$ can be
considered as Chern-Simons theories with the gauge group $SL(N)$. 
Moreover, for even $N=2n$ such models admit a truncation to the fields
with even spins $2,4,\dots,2n$ only with corresponding group being
$Sp(2n)$ \cite{Vas91,CLW12}. The simplest representative of these
models is the one with group $Sp(4)$ containing just two fields with
spin-2 and spin-4. But there exists a very well-known (but mostly in
four dimensions) supergroup $OSp(1,4)$ that appears as the natural
supersymmetric extension of the $Sp(4)$. Moreover, using a so-called
principal embedding of $Sp(2)$ (see e.g. explicit expressions in
Appendix A of \cite{CLW12}) it follows that corresponding model must
contain fermionic field with spin 5/2. Here we give a direct
construction of such model.

We will use local $Sp(4)$ indices $a,b=1,2,3,4$ which will be raised
and lowered with antisymmetric $Sp(4)$ invariant tensor $E^{ab}$
normalized so that $E^{ab} E_{bc} = - \delta^a{}_c$. Then we introduce
bosonic one-form $\Omega^{ab}$ and fermionic one-form $\Psi^a$. Note
now that all calculations in the previous section demonstrating gauge
invariance of the Lagrangian will not change if one will replace
spinor indices $\alpha,\beta$ by $a,b$ and $\varepsilon^{\alpha\beta}$
by $E^{ab}$. Thus we immediately obtain the Lagrangian
\begin{eqnarray}
{\cal L} &=& \frac{1}{2} \Omega_{ab} d \Omega^{ab} + \frac{i}{2}
\Psi_a d \Psi^a \nonumber \\
 && + \frac{\tilde{a}_0}{3} \Omega_a{}^b \Omega_b{}^c \Omega_c{}^a -
\frac{i\tilde{a}_0}{2} \Psi_a \Omega^{ab} \Psi_b
\end{eqnarray}
where $\tilde{a}_0 = \sqrt{10}a_0$ as well as corresponding gauge
transformations:
\begin{eqnarray}
\delta \Omega^{ab} &=& d \eta^{ab} + \tilde{a}_0 \Omega^{(a}{}_c
\eta^{b)c} + \frac{i\tilde{a}_0}{2} \Psi^{(a} \xi^{b)} \nonumber \\
\delta \Psi^a &=& d \xi^a + \tilde{a}_0 \eta^{ab} \Psi_b - \tilde{a}_0
\Omega^{ab} \xi_b
\end{eqnarray}
Equations for this Lagrangian look like:
\begin{eqnarray}
 && d \Omega^{ab} + \tilde{a}_0 \Omega^a{}_c \Omega^{bc} + 
\frac{i\tilde{a}_0}{2} \Psi^a \Psi^b = 0 \nonumber \\
 && d \Psi^a + i\tilde{a}_0 \Omega^a{}_b \Psi^b = 0
\end{eqnarray}
Now let us switch back to the multispinor formalism by the rule $a
\Rightarrow (\alpha_1\alpha_2\alpha_3)=\alpha(3)$. Thus we introduce:
\begin{eqnarray*}
\Omega^{ab} &\Rightarrow& \Omega^{(\alpha(3),\beta(3))} =
\Sigma^{\alpha(3)\beta(3)} + \frac{1}{6\sqrt{10}}
\varepsilon^{\alpha\beta} \varepsilon^{\alpha\beta}
\Omega^{\alpha\beta} \\
\Psi^a &\Rightarrow& \Psi^{\alpha(3)}
\end{eqnarray*}
where $\Sigma^{\alpha(6)}$ is completely symmetric six-spinor
corresponding to spin-4 field, while two other fields correspond to
spin-2 and spin-$\frac{5}{2}$. Here and in what follows the complete
symmetrization over all spinor indices denoted by the same letter is
assumed. Equations in terms of new variables take the form:
\begin{eqnarray}
 && d \Omega^{\alpha\beta} + a_0 \Omega^\alpha{}_\gamma
\Omega^{\beta\gamma} + 3a_0 \Sigma^\alpha{}_{\gamma(5)} 
\Sigma^{\beta\gamma(5)} + \frac{3ia_0}{2} \Psi^\alpha{}_{\gamma(2)}
\Psi^{\beta\gamma(2)} = 0 \nonumber \\
 && d \Sigma^{\alpha(6)} + a_0 \Omega^{\alpha}{}_\beta
\Sigma^{\alpha(5)\beta} + \frac{\tilde{a}_0}{2} 
\Sigma^{\alpha(3)}{}_{\beta(3)} \Sigma^{\alpha(3)\beta(3)} +
\frac{i\tilde{a}_0}{20} \Psi^{\alpha(3)}
\Psi^{\alpha(3)} = 0 \\
 && d \Psi^{\alpha(3)} + a_0 \Omega^\alpha{}_\beta
\Psi^{\alpha(2)\beta} + \tilde{a}_0
\Sigma^{\alpha(3)}{}_{\beta(3)} \Psi^{\beta(3)} = 0 \nonumber
\end{eqnarray}
It is easy to see that we still have $AdS_3$ background as a solution
$$
\Omega_0{}^{\alpha\beta} = \frac{1}{a_0} \hat{\omega}^{\alpha\beta},
\qquad \Sigma_0{}^{\alpha(6)} = 0, \qquad \Psi_0{}^{\alpha(3)} = 0
$$
So for the $AdS_3$ background we obtain
$$
{\cal L} = {\cal L}_0 + {\cal L}_{1b} + {\cal L}_{1f}
$$
\begin{eqnarray}
{\cal L}_0 &=& \frac{1}{2} \omega_{\alpha\alpha} D
\omega^{\alpha\alpha} + \frac{\lambda}{2}
\omega_{\alpha\beta} e^\alpha{}_\gamma \omega^{\beta\gamma} \nonumber
\\
 && + \frac{1}{2} \Sigma_{\alpha(6)} D \Sigma^{\alpha(6)} + 
\frac{3\lambda}{2} \Sigma_{\alpha(5)\beta} e^\beta{}_\gamma
\Sigma^{\alpha(5)\gamma} \nonumber \\
&& + \frac{i}{2} \Psi_{\alpha(3)} D \Psi^{\alpha(3)} +
\frac{3i\lambda}{4} \Psi_{\alpha(2)\beta} e^\beta{}_\gamma
\Psi^{\alpha(2)\gamma} \\
 {\cal L}_{1b} &=& \frac{a_0}{3} \omega_\alpha{}^\beta
\omega_\beta{}^\gamma \omega_\gamma{}^\alpha + 3a_0
\Sigma_{\alpha(5)\beta} \omega^\beta{}_\gamma \Sigma^{\alpha(5)\gamma}
\nonumber \\
 && + \frac{10\tilde{a}_0}{3} \Sigma_{\alpha(3)\beta(3)}
\Sigma^{\beta(3)}{}_{\gamma(3)} \Sigma^{\alpha(3)\gamma(3)} \\
{\cal L}_{1f} &=& \frac{3ia_0}{2} \Psi_{\alpha(2)\beta}
\omega^\beta{}_\gamma \Psi^{\alpha(2)\gamma} + 
\frac{3i\tilde{a}_0}{2} \Psi_{\alpha(3)} 
\Sigma^{\alpha(3)}{}_{\beta(3)} \Psi^{\beta(3)} 
\end{eqnarray}
Here ${\cal L}_0$ is a sum of free Lagrangians for massless spin-2,
spin-4 and spin-$\frac{5}{2}$ fields, ${\cal L}_{1b}$ describes
self-interaction for graviton, its interaction with spin-4 and
self-interaction for spin-4 field, while ${\cal L}_{1f}$ introduces
gravitational interaction for spin-$\frac{5}{2}$ field and its
interaction with spin-4. At the same time corresponding gauge
transformations take the form:
\begin{eqnarray}
\delta \omega^{\alpha\alpha} &=& D \eta^{\alpha\alpha} +
\frac{\lambda}{2} e^\alpha{}_\beta \eta^{\alpha\beta} + a_0
\omega^\alpha{}_\beta \eta^{\alpha\beta} \nonumber \\
 && + 3a_0 \Omega_{\beta(5)}{}^\alpha \eta^{\alpha\beta(5)} +
\frac{3ia_0}{2} \Psi_{\beta(2)}{}^\alpha \xi^{\alpha\beta(2)}
\nonumber \\
\delta \Omega^{\alpha(6)} &=& D \eta^{\alpha(6)} + \frac{\lambda}{2}
e^\alpha{}_\beta \eta^{\alpha(5)\beta} \nonumber \\
 && + a_0 \Omega^{\alpha(5)}{}_\beta \eta^{\alpha\beta} + a_0
\omega^\alpha{}_\beta \eta^{\alpha(5)\beta}  \\
 && + \tilde{a}_0 \Omega^{\alpha(3)}{}_{\beta(3)}
\eta^{\alpha(3)\beta(3)} +  \frac{i\tilde{a}_0}{20} \Psi^{\alpha(3)}
\xi^{\alpha(3)} \nonumber \\
\delta \Psi^{\alpha(3)} &=& D \xi^{\alpha(3)} + \frac{\lambda}{2}
e^\alpha{}_\beta \xi^{\alpha(2)\beta} \nonumber \\
 && + a_0 \Psi^{\alpha(2)}{}_\beta \eta^{\alpha\beta} + \tilde{a}_0
\Psi_{\beta(3)} \eta^{\alpha(3)\beta(3)} \nonumber \\
 && + a_0 \omega^\alpha{}_\beta \xi^{\alpha(2)\beta} + \tilde{a}_0
\Omega^{\alpha(3)}{}_{\beta(3)} \xi^{\beta(3)} \nonumber
\end{eqnarray}

Note that there exists a straightforward generalization of this model
to the case of $OSp(1,2n)$ supergroup. Using again principal embedding
(see e.g. Appendix C in \cite{CLW12}) we obtain model containing a
number of bosonic fields with even spins $2,4,\dots,2n$ and one
fermionic field with spin $n+1/2$.

\section*{Conclusion}

One of the important properties of three-dimensional higher spin
theories is that there exist not only models with infinite number of
higher spin fields as e.g. in \cite{Ble89,PV98,HGPR12}, but also
models with finite number of them, see e.g.
\cite{Vas91,CFPT10,CLW12,Tan12}. In this paper we have presented a
whole class of models containing a number of bosonic fields with spins
$2,4,\dots 2n$ and one fermionic field with spin $n+\frac{1}{2}$. As
far as we know these are the simplest models containing fermions and
as we have shown they can be considered as a straightforward
generalization of the minimal $(1,0)$ supergravity. Taking into
account this similarity with the supergravity it may be tempting to
try to construct some kind of matter hypermultiplets and consider
their interaction with hypergravity. But the results of
\cite{Vas91,PV98} show that such a program is hardly possible.
Certainly there have to exist similar generalization for extended
supergravities but exploring this possibility lies beyond the scope of
the current paper.

\section*{Acknowledgments}
Author is grateful to T.~V.~Snegirev for collaboration. Work was
partially supported by RFBR grant No. 14-02-01172.

\end{document}